\documentclass[aps,pre,reprint,nofootinbib]{revtex4-2}
\usepackage{blindtext}
\usepackage{graphicx}
\usepackage{dcolumn}
\usepackage{bm}
\usepackage{mathtools, nccmath}
\usepackage{xcolor}
\usepackage[hidelinks]{hyperref}
\newcommand{\rev}[1]{#1}
\setcitestyle{super,comma}
\setcitestyle{super,comma,sort&compress}   

\makeatletter
\renewcommand\@biblabel[1]{#1}             
\makeatother
\providecommand{\bibnumfmt}[1]{#1}
\renewcommand{\bibnumfmt}[1]{#1}

\renewcommand{\figurename}{Fig.}           

\begin{document}
\makeatletter
\renewcommand{\figurename}{Fig.}
\def\@caption@fignum@sep{ }
\makeatother
\title{Effect of Small Particle Interactions on Assembly of Binary Colloidal Clusters}

\author{Dillip Kumar Mohapatra\textsuperscript{*}}
\author{Teun W.J. Verouden}
\author{Bos Everhardus Sprik}
\author{Wouter G. Ellenbroek}
\author{Janne-Mieke Meijer\textsuperscript{*}}

\affiliation{Department of Applied Physics and Science Education, Eindhoven University of Technology, Groene Loper 19, 5612 AP Eindhoven, The Netherlands.}
\affiliation{Institute for Complex Molecular Systems, Eindhoven University of Technology, Groene Loper 19, 5612 AP Eindhoven, The Netherlands.}




\begin{abstract}
A central challenge in colloidal self-assembly is understanding how interaction strength influences the competition between structural optimization and kinetic arrest during the formation of hierarchical structures. Here, we investigate the assembly process of a binary mixture of differently sized silica colloids dispersed in a near-critical water/2,6-lutidine solvent, where critical Casimir forces provide reversible and temperature tunable attractions. By systematically tuning the effective interactions between large-large (L-L), large-small (L-S), and small-small (S-S) particle pairs, we directly observe how the nucleation and growth of the binary colloid assemblies evolve \rev{in a wall-confined quasi-two-dimensional geometry} when multiple pair-interactions are present.
Confocal laser scanning microscopy imaging shows that combined L-L interactions and L-S interactions lead to compact L-S core-shell clusters with high shell occupancy, while the additional presence of S-S interactions suppress the shell formation and lead to extended, branched binary networks. 
Particle-resolved analyses of cluster growth and bond dynamics reveal that these structural transitions are governed by the motion of the small particles in the clusters changing from highly dynamic leading to structural reorganization to frozen dynamics resulting kinetic trapping. 
Our findings demonstrate how the interaction-governed mobility and kinetic arrest of small particles influence the growth of finite binary clusters and offers microscopic insight into the formation of binary crystals and kinetically trapped space-spanning binary colloidal networks and gels.
\end{abstract}

\maketitle
{\renewcommand{\thefootnote}{}%
\footnotetext{\textsuperscript{*}d.k.mohapatra@tue.nl, j.m.meijer@tue.nl; \textsuperscript{*}Corresponding authors.}}
\section{Introduction}

Understanding how microscopic interactions give rise to the formation of complex hierarchical superstructures remains a central challenge in soft matter and materials science. 
Extensive studies on colloidal systems have shown that complex particle interactions, dynamics, and assembly pathways control the formation of crystals, glasses and gels \cite{manoharan_colloidal_2015, boles_self-assembly_2016, royall_real_2021}.
When combining colloidal spheres of two different sizes, also known as a binary colloid mixture, the phase behavior becomes even richer, arising from the interplay between particle size asymmetry, composition, diffusivity, and interparticle interactions \cite{diba_binary_2018,boles_self-assembly_2016}. 
Binary systems have been shown to form a wide range of equilibrium and non-equilibrium structures, including binary crystals\cite{huang_binary_2007, khalil_binary_2012, immink_crystal--crystal_2020, raybin_nonadditive_2023}, colloidal alloys \cite{ducrot_colloidal_2017, fernandez-rodriguez_tunable_2018}, binary gels\cite{immink_reversible_2019}, binary glasses \cite{nie_connecting_2020} and colloidal molecules \cite{erb_magnetic_2009, schade_tetrahedral_2013, heatley_colloidal_2017, mu_binary_2022, chakraborty_self-assembly_2022}. These binary materials are of interest because these can possess unique optical\cite{yang_multidimensional_2025} and electronic properties \cite{lopez-godoy_self-assembly_2022}. 
For these systems, the strength of the interactions between the particles not only drives the particle association but also influences the assembly kinetics \cite{ghofraniha_assembly_2009} and the ability of particles to assemble and reorganize during the assembly \cite{chakraborty_self-assembly_2022}. For instance, when in a binary colloid mixture one particle size will assemble/aggregate before the other, interaction-driven fractionation has been reported \cite{yao_synchronized_2020} and intertwined or core-shell gel networks with different mechanical properties \cite{immink_reversible_2019}. Also in binary mixtures of active and passive particle colloidal molecules and fractionation have been reported \cite{wang_active_2020, schmidt_light-controlled_2019}. 
Clearly, the final hierarchical structure that forms depends on details of the interparticle interactions but also on the assembly pathways that reflect a competition between structural optimization and kinetic arrest.

So far, most binary colloid assemblies have been realized using a range of methods that can achieve specific interactions between the large and small particles. These interactions are extensive and include DNA-mediated binding \cite{fang_two-step_2020, chakraborty_self-assembly_2022}, biotin-streptavidin bonding \cite{ghofraniha_assembly_2009}, depletion attractions \cite{waheibi_pairing-specific_2024, tahmid_shahriar_frustrated_2025}, electrostatic interactions \cite{yao_synchronized_2020, leunissen_ionic_2005}, magnetic fields \cite{khalil_binary_2012}, solvent evaporation techniques \cite{dai_phase_2012, velikov_layer-by-layer_2002} or liquid confinement \cite{guo_programmed_2018}. Although these different interactions have enabled the fabrication of diverse structures, they generally provide limited dynamic control after the attractive interactions are induced and aggregation begins. Consequently, experiments often probe only the final assembled states, while the intermediate stages of nucleation, growth, restructuring, and kinetic arrest are only limitedly explored \cite{ghofraniha_assembly_2009}. As a result, the microscopic mechanisms linking particle dynamics to assembly outcomes remain challenging to determine.

To fully understand the emergence of binary colloid assemblies, it is essential to track the entire assembly process at the single-particle level. Although simulations can provide microscopic insight, they often cannot capture the long timescales and non-equilibrium pathways characteristic of colloidal assembly \cite{wang_kinetically_2020, zhou_discovery_2022, lacour_influence_2019, mao_deep_2022, tasios_critical_2016}. Micron-sized colloids offer a unique opportunity to investigate the assembly process of binary colloid system because their Brownian motion and assembly dynamics can be directly visualized using optical microscopy \cite{poon_colloids_2004}. In addition, recently critical Casimir interactions have been shown to be tunable, reversible, and sufficiently gentle to provide access to in-situ assembly and thereby allow imaging of intermediate assembly states\cite{bonn_direct_2009, hertlein_direct_2008} . The critical Casimir interactions is a solvent-mediated force that arises in colloidal suspensions dispersed in a near-critical binary liquid mixture that phase separates upon heating. 
By approaching the solvent critical temperature, the correlation length of the solvent fluctuations increases, and leads to effective attractive interactions between colloidal particles composed of the same material \cite{stuij_critical_2017}. The strength of the attractive interactions between colloids depends on the solvent correlation length and the overlap volume between two colloidal spheres \cite{bonn_direct_2009}. 
This control enables continuous and reversible tuning of the interaction strength through small temperature increments  \cite{hertlein_direct_2008, gambassi_critical_2009}. Previous studies have demonstrated the versatility of the critical Casimir interactions in directing the assembly of monodisperse colloids \cite{hertlein_direct_2008, rouwhorst_nonequilibrium_2020, nguyen_controlling_2013}, anisotropic particles \cite{kennedy_self-assembly_2022}, and patchy building blocks \cite{swinkels_networks_2024, stuij_revealing_2021,swinkels_visualizing_2023}. 

Here, we employ the critical Casimir force to induce the assembly of a binary mixture of differently sized silica colloids into \rev{finite-sized} colloidal clusters in the presence of competing pair-interactions. We achieve this by dispersing the colloids in a near-critical water/2,6-lutidine (W/L) mixture and using a temperature-controlled confocal laser scanning microscope (CLSM) to directly visualize the assembly process \cite{nguyen_critical_2016,kennedy_self-assembly_2022}. 
By tuning the temperature we systematically vary the relative strengths of the interactions between large-large (L-L), large-small (L-S), and small-small (S-S) particle pairs. We investigate the self-assembly, formation and growth of different colloidal clusters ranging from L-S core-shell clusters to large branched binary networks. We further investigate how the strength of the L-S and S-S interactions influence the small particle dynamics during the assembly of the colloidal clusters. Our work experimentally demonstrates that the small particle interactions and dynamics crucially govern the growth of finite binary colloidal clusters and provides insight into the mechanisms that cause kinetically trapping in system-spanning binary colloidal networks.

\section{Experimental}

\subsection{Binary colloid system}
Two differently sized silica spheres were purchased from Microparticles GmbH and used without further surface modification. The particle SiO2-FluoGreen-4.0 has a diameter of $3.87~\pm~0.12~\mu\mathrm{m}$ (Large) and particle SiO2-FluoOrange-1.0 has a diameter of $0.966~\pm~0.04~\mu\mathrm{m}$ (Small). Both particles are labeled with a fluorescent dye and were dispersed in water at concentrations of $2.5~wt\%$.
This size ratio was chosen to achieve a large enough temperature difference between the temperatures where the strength of the critical Casimir interaction of the pairs of large–large (L-L), large–small (L-S), and small–small (S-S) becomes of the order of the thermal energy $\simeq k_B T$ to be experimentally accessible. This was determined by calculating the total interaction potential \(V_{total}(l)\) between two spherical particles, according to \cite{bonn_direct_2009}:
\begin{equation}
    V_{{total}}(l) = V_{\mathrm{Electrostatic}}(l) + V_{\mathrm{Casimir}}(l) 
\end{equation}
\begin{equation}
 V_{total}(l) \approx \frac{4\pi R_1 R_2 \sigma^2 \lambda_d^2} {\epsilon \epsilon_0 (R_1 + R_2)} e^{-l/\lambda_d} - \frac{4\pi R_1 R_2 k_B T} {\xi (R_1 + R_2)} e^{-l/\xi} 
 \end{equation}
 \rev{
\begin{equation}
V_{{total}}(l)
\approx
4\pi R_{\mathrm{eff}}
\left[
\frac{\sigma^2\lambda_d^2}{\epsilon\epsilon_0}
e^{-l/\lambda_d}
-
\frac{k_B T}{\xi}
e^{-l/\xi}
\right]
\end{equation}
}
where, \rev{$R_{\mathrm{eff}}=R_1R_2/(R_1+R_2)$ is the effective radius of the interacting particle pair, with $R_1$ and $R_2$ denoting the radii of the two particles,} $l$ is the surface-to-surface distance between two colloids, $\sigma$ is the surface charge density, $\lambda_d$ is the Debye screening length, $\epsilon$ is the dielectric constant of the solvent, and $\epsilon_0$ is the vacuum permittivity, $k_B$ is the Boltzmann constant, $T$ is the temperature. \rev{The bulk correlation length, $\xi$, is defined as $\xi = \xi_0 \left(1-\frac{T}{T_c}\right)^{-\nu}$, where $T_c$ is the critical temperature of the W/L mixture, $\xi_0$ is the correlation-length amplitude reflecting the characteristic length scale set by the intermolecular pair potential, and $\nu = 0.63$ is the universal critical exponent \cite{hertlein_direct_2008}.}
We note that due to difficult nature of measuring $\sigma$ of the silica colloids and the value of  $\lambda_d$ in the W/L mixture, the calculation is not exact but did provide a good estimate for the temperature window in which the different interactions, L-L, L-S and S-S are expected to become significantly strong to induce assembly, as schematically shown in Fig. ~\ref{fig:Schematics_Expt_Setup}.
\rev{To further characterize the particle surface properties, we measured the zeta potential of both particle sizes in water and in the W/L mixture (Fig. S1). Both particle remain strongly negatively charged, with measured zeta potentials of approximately $-58$~mV and $-61$~mV in water for the $1~\mu\mathrm{m}$ and $4~\mu\mathrm{m}$ spheres, respectively. In the W/L mixture, the corresponding zeta potentials are approximately $-156$~mV and $-170$~mV, respectively. No salt was added to the samples. Since the ionic strength of the mixed solvent was not independently measured, we do not assign an exact Debye screening length.
To provide an estimate of the interaction-energy scale, we additionally estimated the effective S-S interaction from the measured radial distribution function using \(U_{SS}(r) = -k_{\mathrm{B}}T \ln g_{SS}(r)\) \cite{mohanty_direct_2026} as shown in Supplementary Fig.~S2. 
At $\Delta T= T_c -T = 0.55\,^{\circ}\mathrm{C}$, this gives a potential minimum of approximately $-1\,k_{\mathrm{B}}T$. Using the effective-radius dependence in Eq.~(3), the corresponding interaction scales are estimated to be approximately $-1.6\,k_{\mathrm{B}}T$ for L-S and $-4\,k_{\mathrm{B}}T$ for L-L interactions at the same temperature, as shown in Fig.~\ref{fig:Schematics_Expt_Setup}(b). These values are estimates of the interaction scale rather than independently measured L-S and L-L pair potentials. }
 
Samples were prepared as follows. First, appropriate amounts of each particle dispersion were transferred into Eppendorf tubes and washed three times with a near-critical 2,6-lutidine (\(\geq 99.0~\%\), Sigma-Aldrich)/water (Milli-Q) solvent mixture (28.4~wt\% lutidine) by centrifugation at $1000~\mathrm{rcf}$ for $5~\mathrm{min}$. Next, the overall particle concentration was adjusted by adding the required amount of the W/L mixture to obtain a total area fraction of $\phi_A \approx$ 0.08 and a particle number ratio (large:small = 1:12). This particle number ratio was selected based on geometric packing considerations. 
The maximum number of small colloids surrounding a central colloid in \rev{the wall-confined geometry} can be estimated geometrically as $N_{\mathrm{max}} = \pi / \arcsin\!\left(\dfrac{1}{2}\sqrt{\dfrac{R_o}{R_i}}\right)$, where $R_i$ the radius of the inner colloid (on the contact plane) and $R_o$ the radius of the outer colloids. \rev{Here, both particles can contact the substrate, while their centers lie at different heights because of their different radii. The resulting wall-confined geometry gives a maximum first-shell occupancy of approximately 12 small particles} (See Supplementary Information for details). 
Next, the dispersion was placed in rectangular borosilicate capillaries (0.2~$\times$~2~$\times$~50~mm, VitroCom) with hydrophilic inner surfaces \cite{kennedy_self-assembly_2022}. \rev{Prior to use, the capillaries were soaked with 100 times diluted Hellmanex III solution for 2 hours to increase hydrophilicity of the glass and remove impurities. Then they were cleaned with milliQ water and dried of with nitrogen gas. After loading the dispersion, both ends of the capillaries were sealed with} Teflon grease (Krytox). 
\rev{In addition to particle-particle interactions, critical Casimir interactions with the capillary wall can contribute to the confinement. From the effective-radius dependence in Eq.~(3), the particle-wall interaction is estimated to be approximately twice as strong as the corresponding interaction between two identical particles at the same surface separation and temperature, assuming comparable surface interaction parameters (Fig.~\ref{fig:Schematics_Expt_Setup}(b)). This is an estimate because the surface properties of the treated capillary wall were not independently measured.}

\subsection{Temperature-controlled microscopy}

\begin{figure}
    \centering
    \includegraphics[width=\linewidth]{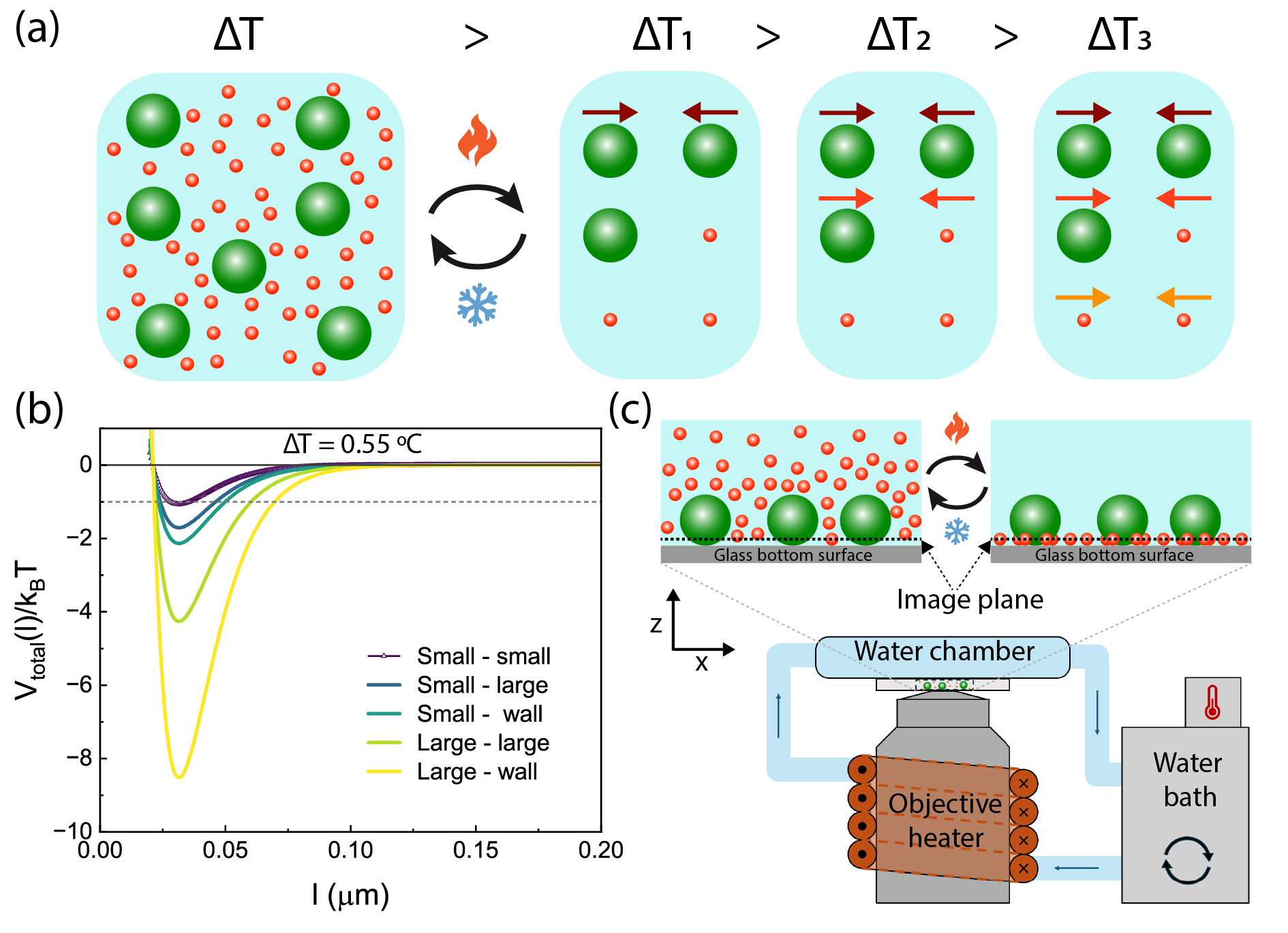}
    \caption{ (a) Schematic illustration of the temperature control over the strength of the interactions between the two differently sized silica colloidal spheres in the critical W/L mixture as the temperature approaches the critical temperature where $\Delta T = 0$. For large $\Delta T$ no interactions are present between all colloids, for decreasing $\Delta T$,  attractive interactions between large-large, large-small, and small-small particle pairs emerge at distinct different $\Delta T$. Arrows pointing toward each other ($\rightarrow\!~\leftarrow$) represent the presence of attractive interparticle interactions. 
    \rev{(b) Effective interaction potentials, $V_{\mathrm{total}}(l)/k_{\mathrm{B}}T$, calculated at $\Delta T = 0.55\,^{\circ}\mathrm{C}$ for small-small, small-large, small-wall, large-large, and large-wall pairs. The effective S–S interaction strength was estimated experimentally from $g_{SS}(r)$ at $\Delta T=0.55~^\circ\mathrm{C}$, while the interaction strengths for the other pairs were estimated using the effective-radius scaling in Eq.~(3). The different potential depths illustrate the size-dependent strength of the attractive interactions.
    (c) Schematic of the temperature-controlled confocal microscopy setup. The enlarged view shows the wall-confined imaging geometry near the bottom capillary surface, with the focal plane positioned approximately at the mid-plane of the small particles ($z \sim 0.5\,\mu\mathrm{m}$ above the wall)}}
    \label{fig:Schematics_Expt_Setup}
\end{figure}

Sample capillaries were mounted on a Nikon ECLIPSE Ti2e inverted microscope equipped with a AR1-25HD scan head and a custom-built temperature-controlled sample stage setup (OKO-lab) capable of maintaining a thermal stability of $0.02^\circ\mathrm{C}$.
The temperature control system consist of a water-circulating bath (Lauda RE 415 s), an objective heater (metal collar surrounding the objective), and a glass heating chamber, as also schematically shown in Fig.~\ref{fig:Schematics_Expt_Setup}(c).
Water was circulated through both the objective heater and the chamber to ensure the objective lens (100x), immersion oil, chamber, and sample capillary equilibrated to the set temperature.
Fluorescence imaging was performed using 488~nm (BODIPY) and 561~nm (Rhodamine Red~X) laser excitation to selectively visualize the large and small particles, respectively. 
\rev{Images were acquired close to the bottom capillary wall, with the focal plane positioned approximately at the mid-plane of the small particles (\(z\sim0.5~\mu{\rm m}\) above the wall), rather than at the mid-plane of the large particles.}

Assembly experiments were performed over a range of temperatures, according to the following protocol. The samples were first equilibrated at $33^\circ\mathrm{C}$, a temperature at which no critical Casimir attractions are known to occur between the colloids. At this temperature, we observed that all large particles sedimented onto the capillary wall due to the density mismatch between the silica and the solvent and their size. In contrast, the small silica colloids were found to remain largely suspended over the experimental timescale in the absence of attractive interactions. 
To trigger the self-assembly process, the sample temperature was rapidly increased to a desired temperature.  For the temperatures where L-S and S-S interactions are present, we observed that the small colloids also settled on the wall. 
\rev{The assembly therefore occurs in a wall-confined quasi-2D geometry. Sedimentation and particle-wall interactions restrict out-of-plane motion and influence particle mobility and consequently the assembly kinetics.}
Time-series of 2D images were recorded of a field of view 174 × 87 $\mu$m with a resolution of 1024 × 512 pixels and a frame rate of 0.5~frame/s for $30$ min. After each measurement, the sample temperature was reduced again to $33^\circ\mathrm{C}$ and held for 10 min to allow the system to return to its initial dispersed state, after which the next measurement was performed.
\rev{The sample was not mechanically stirred between consecutive measurements. Before starting the subsequent measurement, we visually confirmed by microscopy that no persistent clusters remained.}
Because the solvent critical temperature $T_c$ is highly sensitive to solvent composition, we determine $T_c$ \textit{in-situ} as the point where macroscopic phase separation of the W/L mixture occurs. For this, after completing all measurements, the sample temperature is slowly increased until the phase separation is observed. For the solvent composition used here, we measured $T_c = 34^\circ\mathrm{C}$. All temperatures are reported relative to the critical temperature according to $\Delta T = T_c - T$ and range between  $\Delta T = 0.65~^\circ\mathrm{C}$ and $\Delta T = 0.25~^\circ\mathrm{C}$.
\rev{Over this temperature range, the estimated solvent correlation length increases from approximately $10$ to $18\,\mathrm{nm}$.
To quantify the effect of confinement on particle mobility, we additionally measured the diffusion coefficients of both particles in dilute samples (\(\phi_A=0.02\)) as a function of \(\Delta T\); the results are shown in Fig. S3.}

\subsection{Particle tracking}
Particle positions and trajectories were obtained from the CLSM data using the centroid-based tracking algorithm TrackPY based on the Crocker-Grier approach \cite{crocker_methods_1996, allan_soft-mattertrackpy_2021}. 
Particles in the same cluster were identified via nearest-neighbor pairs. The pairs were identified using distance cutoffs of 4.25~$\mu$m (25 pixels) for large-large, 2.5~$\mu$m (15 pixels) for large-small, and 1.19~$\mu$m (7 pixels) for small-small particle pairs. The L-L and S-S cutoffs correspond to approximately 1.1 and 1.2 times the particle diameters, respectively. For L-S pairs, the cutoff was chosen as the contact distance between the large and small particles, as we observed that a larger cutoff led to occasional inclusion of second-nearest neighbors due to the size asymmetry of the system. 
\rev{A small particle was classified as free when no large particle was located within the L-S cutoff distance of \(2.5~\mu{\rm m}\) and no other small particle was located within the S-S cutoff distance of \(1.19~\mu{\rm m}\).
The bond angles of small satellite particles around a large particle were calculated from their tracked positions. To examine angular rearrangements within the first coordination shell, the angular positions of individual satellite particles were tracked relative to a selected reference satellite, whose relative angle was set to $0^\circ$ at each time point.}


\section{Results and Discussion}

\subsection{Assembly of binary colloid clusters}

\begin{figure*}
\centering
    \includegraphics[width=\textwidth]{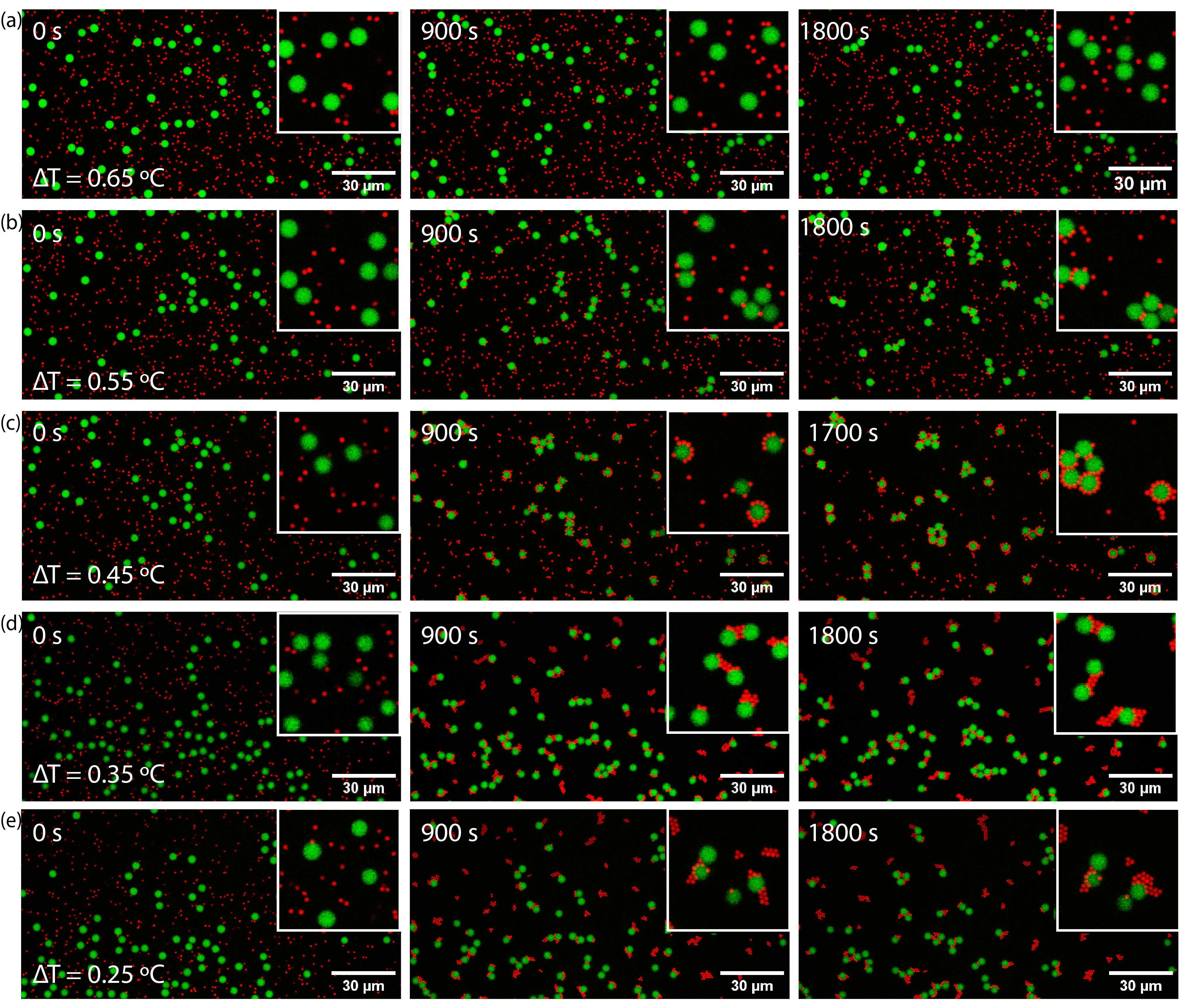}
    \caption{
    (a - e) CLSM images at different time steps during the assembly process of \rev{finite-sized clusters} in the binary system of differently sized colloids, large (green) and small (red), at a area fraction of \(\phi_A = 0.08\) at different temperatures  \(\Delta T\) = \(0.65, 0.55, 0.45, 0.35\) and \(0.25^\circ\mathrm{C}\) that controls the strength of the attractive interactions between the large-large, large-small and small-small particle pairs. The large to small number ratio is 1:12 that should facilitate full first-shell coverage of the large particles by the small particles in \rev{the wall-confined geometry}. (Insets) Enlarged views of formed binary clusters.}
    \label{fig:ConfImgTemporalEvolution} 
\end{figure*}

Since the strength of the temperature-tunable critical Casimir interactions depends on the solvent correlation length and the overlap volume around a colloid \cite{bonn_direct_2009,hertlein_direct_2008}, in the binary colloid mixture of large ($D_L= $$3.87~\pm~0.12~\mu\mathrm{m}$ ) and small ($D_S = 0.966~\pm~0.04~\mu\mathrm{m}$) we can control the presence of attractive large-large, large-small and small-small colloid pair-interactions with temperature. To investigate how the presence of multiple pair-interactions influence the assembly of the binary colloid system, we systematically varied the strength of the pair-interactions by tuning $\Delta T$, i.e. the distance from the solvent critical temperature, and followed the formation of clusters and their growth over time with CLSM. Because we are interested in understanding the mechanisms during nucleation and growth of binary assemblies, at each $\Delta T$ we observed the process for a duration of 30 min which is enough time for \rev{finite-sized} clusters to form and grow. After each $\Delta T$  the temperature was lowered to reset the system to a non-interacting fluid. 
Fig.~\ref{fig:ConfImgTemporalEvolution} shows CLSM images of the binary colloid mixture at different time points for each investigated $\Delta T$. 
For the highest investigated $\Delta T = 0.65~^\circ\mathrm{C}$, no attractive interactions between the colloids are observed as expected (Fig.~\ref{fig:ConfImgTemporalEvolution}(a)). Both the large and small particles are homogeneously distributed throughout the field of view, and the suspension exhibits fluid-like behavior without detectable clustering.
Upon decreasing $\Delta T$ to $0.55~^\circ\mathrm{C}$, \rev{weak L–L interactions become evident, whereas L–S and S–S attractions remain too weak to produce persistent association on the experimental timescale} (Fig.~\ref{fig:ConfImgTemporalEvolution}(b)). The formation of \rev{finite-sized} clusters of the large particle is observed, including dimers, trimers, and tetramers, while the small particles remain dispersed . Due to the relatively low diffusivity of the large particles, the growth of the L-L clusters proceeds slowly. 
\rev{The measured large-particle diffusion coefficient decreases from approximately \(0.023-0.024~\mu{\rm m}^2{\rm s}^{-1}\) at larger \(\Delta T\) to approximately \(0.020~\mu{\rm m}^2{\rm s}^{-1}\) closer to \(T_c\) (Fig. S3), confirming increasingly restricted mobility.}
At an intermediate attraction strength, $\Delta T = 0.45~^\circ\mathrm{C}$, the assembly changes markedly and both L-L and L-S pair interactions appear (Fig.~\ref{fig:ConfImgTemporalEvolution}(c) and Supplementary Movie S1). Small particles  accumulate around the large particle clusters. As assembly proceeds, the number of small particles surrounding each large particle increases, resulting in the progressive filling of the first coordination shell and the formation of compact core-shell structures. The shell population continues to increase throughout the experimental window, indicating ongoing structural evolution and reorganization within the clusters. We note that due to a slight decrease in the bath temperature after approximately $1700~\mathrm{s}$, the assembly process was affected beyond this point. Therefore, the final frame shown for $\Delta T = 0.45~^\circ\mathrm{C}$ corresponds to $t = 1700~\mathrm{s}$, unlike the other $\Delta T$ conditions.
For stronger attractions, $\Delta T \leq 0.35~^\circ\mathrm{C}$, all interactions pairs L-L, L-S and S-S appear to be strong and particle attachment occurs rapidly upon meeting and appears irreversible (Fig.~\ref{fig:ConfImgTemporalEvolution}(d,e)) (see Supplementary Movie S2 for $\Delta T = 0.25~^\circ\mathrm{C}$). Small particles quickly bind both to the large particles and to one another, where the latter results in clusters of small particles with local hexagonal symmetry. The final clusters are asymmetric and branched  consisting of large particles connected via \rev{locally hexagonally ordered} domains of small particles. \rev{This local ordering does not necessarily imply long-range crystallization. Rather, local close-packed configurations can form during particle association, while subsequent configurational rearrangements become strongly suppressed at these interaction strengths.} Interestingly, the occupancy of the first coordination shell at this $\Delta T$ exhibits little temporal evolution after the initial attachment events. Instead, newly arriving particles attach to already-bound particles, giving rise to extended satellite-particle networks. As a consequence, shell completion is suppressed and compact core-shell structures are no longer observed. 
Clearly, upon increasing the number of pair-interactions in the binary colloid system the assembly behavior changes and results in distinctly different structures and morphologies of the binary clusters.

\subsection{Temperature controlled pair-interactions}

\begin{figure}
    \centering
    \includegraphics[width=\linewidth]{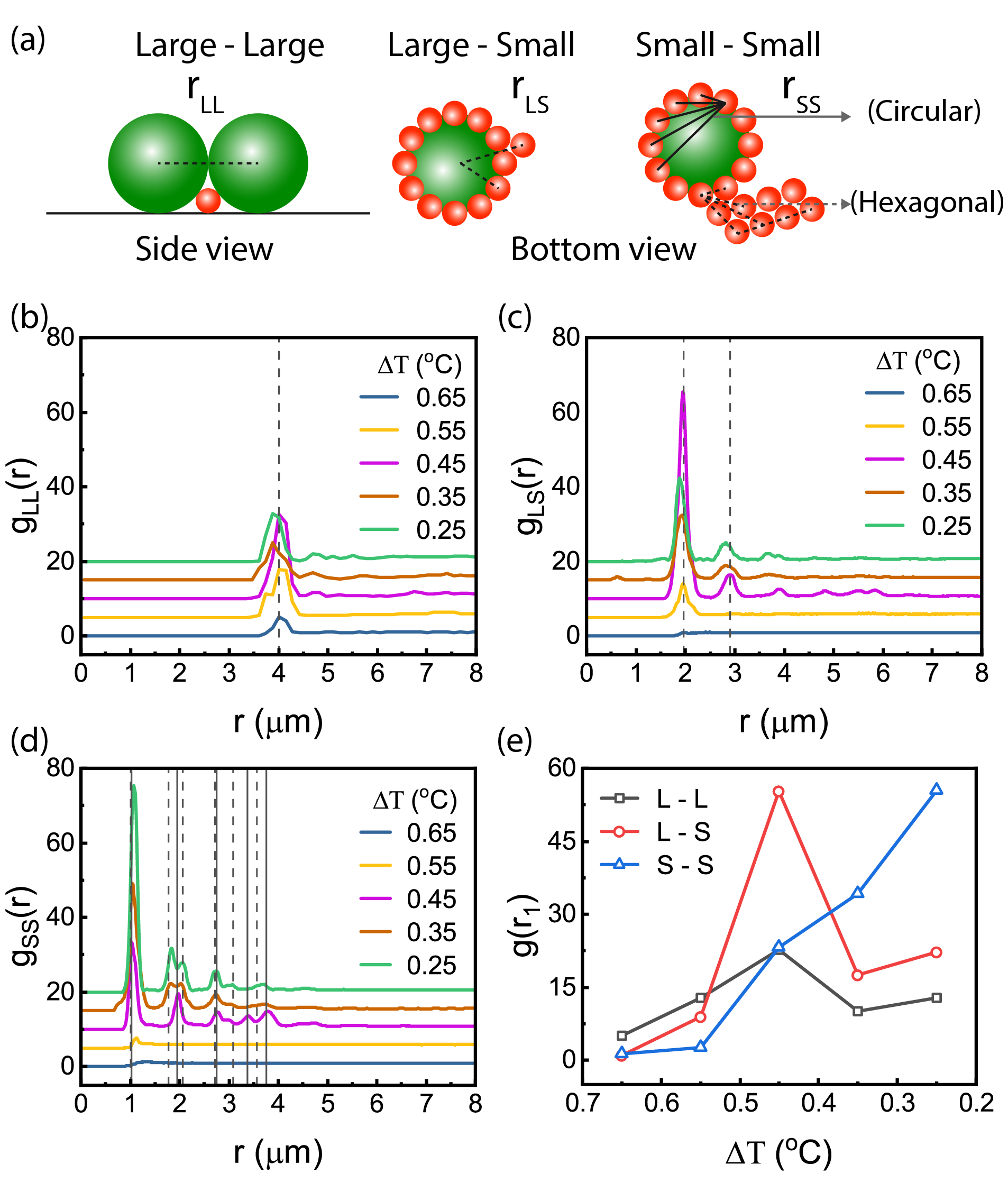}
    \caption{(a) Schematic illustration of characteristic interparticle separations for large-large, large-small, and small-small particle pairs in the binary colloid clusters. For small-small pairs, distances associated with circular (core-shell) and hexagonal arrangements are also indicated. 
    (b-d) Radial distribution functions \(g(r)\) for L-L, L-S, and S-S particle pairs in clusters, respectively, at different attraction strengths \(\Delta T\). Curves are vertically offset by 5 for clarity. Vertical lines indicate the expected interparticle separations shown in (a), with for (d) 
    circular (solid lines) and hexagonal (dashed lines). 
    (e) Height of the first peak \(g(r_1)\) as a function of \(\Delta T\) for all three particle pairs.}
    \label{fig:RDF}
\end{figure}

To verify the temperature control over the presence of the L-L, L-S and S-S pair-interactions and analyze their influence on the structural features of the binary clusters, we examined the pair correlations between the large and small colloids. 
First of all, due to the specific particle size ratio and \rev{the wall-confined quasi-2D geometry resulting from} sedimentation and interactions with the glass capillary wall, the small particles can localize within the confined gap between the wall and the curved surface of the large particles.
\rev{Because the focal plane is positioned approximately at the mid-plane of the small particles, these wall-confined satellite particles can be tracked in the 2D images. The geometric limit \(N_s=12\) therefore corresponds to this wall-confined first coordination shell rather than to an equatorial ring around the large particle.}
Fig.~\ref{fig:RDF}(a) illustrates the configurations of large (green) and small (red) particles on the capillary wall and observed core-shell and branched network configurations, together with the specific interparticle distances. 
For the core-shell configuration, the characteristic distances are given by \(r_n = 2R \sin\left(\frac{n\pi}{N_s}\right)\) where \(r_n\) is the \(n^{th}\) separation, \(R = \frac{D_L}{2} + \frac{D_S}{2}\) is the shell radius, \(D_L\) and \(D_S\) are the diameters of the large and small particles, respectively, and \(N_s\) is the maximum number of small particles surrounding a large particle. 
For the hexagonal arrangement, the interparticle separations are given by \(r_{m,n} = D_S \sqrt{m^2+n^2+mn}\), where \(m,n \in \mathbf{Z}\) are lattice indices. 

Next, we detect all particle positions of the L and S particles and identify particles in a clusters based on distance criteria (see Experimental section).  From the particle positions in all clusters we compute the 2D radial distribution function \(g(r)\) for all L–L, L–S, and S–S particle pairs. 
Fig.~\ref{fig:RDF}(b-d) show all radial distribution function \(g_{LL}(r)\), \(g_{LS}(r)\) and \(g_{SS}(r)\) for different \(\Delta T\). Peaks in the \( g(r)\) correspond to preferred interparticle separations, with the first peak representing the nearest-neighbor shell. The height of the first peak, \(g(r_1)\), quantifies the degree of nearest-neighbor ordering and provides insight into the relative dominance of L-L, L-S, and S-S interactions within the assembled structures.  
In the \(g_{LL}(r)\) (Fig.~\ref{fig:RDF}(b)) a pronounced first peak can be observed for all \(\Delta T \), confirming dominant large-large assembly at weak interaction strengths. The absence of higher-order peaks suggests that the assemblies are primarily limited to dimers or \rev{finite-sized clusters} of large particles, with minimal long-range ordering.
The corresponding \(g_{LS}(r)\) functions (Fig.~\ref{fig:RDF}(c)) shows that for \(\Delta T \leq 0.55~^\circ\mathrm{C}\) distinct peaks appear indicating the emergence of ordering in the clusters. The first-peak position of \(g_{LS}(r)\) is smaller than the nominal center-to-center separation \(R_{LS} = 2.5~\mu\mathrm{m}\), confirming that the small particles occupy the confined gap between the substrate and the curved surface of the large particles. This confinement and interaction strength stabilizes the formation of core-shell structures. The subsequent peaks are the result of S-S correlations arising from the organization of small particles around the large-particle cores in the shell-configuration. 
Fig.~\ref{fig:RDF}(d) shows \(g_{SS}(r)\), where the vertical lines correspond to characteristic interparticle separations for circular shell-like configurations (solid line) and a two-dimensional hexagonal lattice (dashed lines). 
The \(g_{SS}(r)\) curves demonstrate that circular shell-like arrangements dominate at \(\Delta T = 0.45~^\circ\mathrm{C}\), whereas hexagonal ordering of the small particles becomes dominant at \(\Delta T \leq 0.35~^\circ\mathrm{C}\).
\rev{Such local ordering may be facilitated by diffusion and rolling of small particles while they remain bonded, as previously observed during the crystallization of DNA-coated colloids \cite{wang_crystallization_2015}. The characteristic peak positions associated with local hexagonal ordering allow us to distinguish S-S correlations arising from close-packed small-particle domains from those associated with the circular arrangement of satellites around large particles.}
\rev{We note that the S-S interaction was not independently measured in a separate sample containing only small particles. The onset of significant S-S attraction is instead evident from the pronounced S-S correlations and aggregation behavior observed directly in the binary system.}
Fig.~\ref{fig:RDF}(e) summarizes the evolution of the first-peak height \(g(r_1)\) for L-L, L-S, and S-S correlations as a function of \(\Delta T\). Here it is clear that for \(0.45~^\circ\mathrm{C} <\Delta T < 0.55~^\circ\mathrm{C}\) L-L interactions dominate the assembly behavior, for \(\Delta T = 0.45~^\circ\mathrm{C}\) L-S interactions become dominant, leading to the formation of core-shell structures and for \(\Delta T \leq 0.35~^\circ\mathrm{C}\), S-S interactions dominate, resulting in local hexagonal ordering of the small particles. 
Overall, the RDF analysis reveals that decrease $\Delta T$ shifts the dominant pair-interaction from L-L to L-S and finally to S-S and drives a transition in the dominant local packing motif and the preferred mode of organization from shell-like arrangements of small particles to local hexagonal ordering with increasing attraction strength. 
\rev{At the relatively low particle concentration investigated here, the assemblies remain finite over the experimental timescale, whereas higher particle concentrations could favor the formation of extended, fractal-like networks, as commonly observed during colloidal gelation \cite{lu_gelation_2008}.}

\subsection{Small particle growth dynamics and reorganizations}
\begin{figure}
    \centering
    \includegraphics[width=\linewidth]{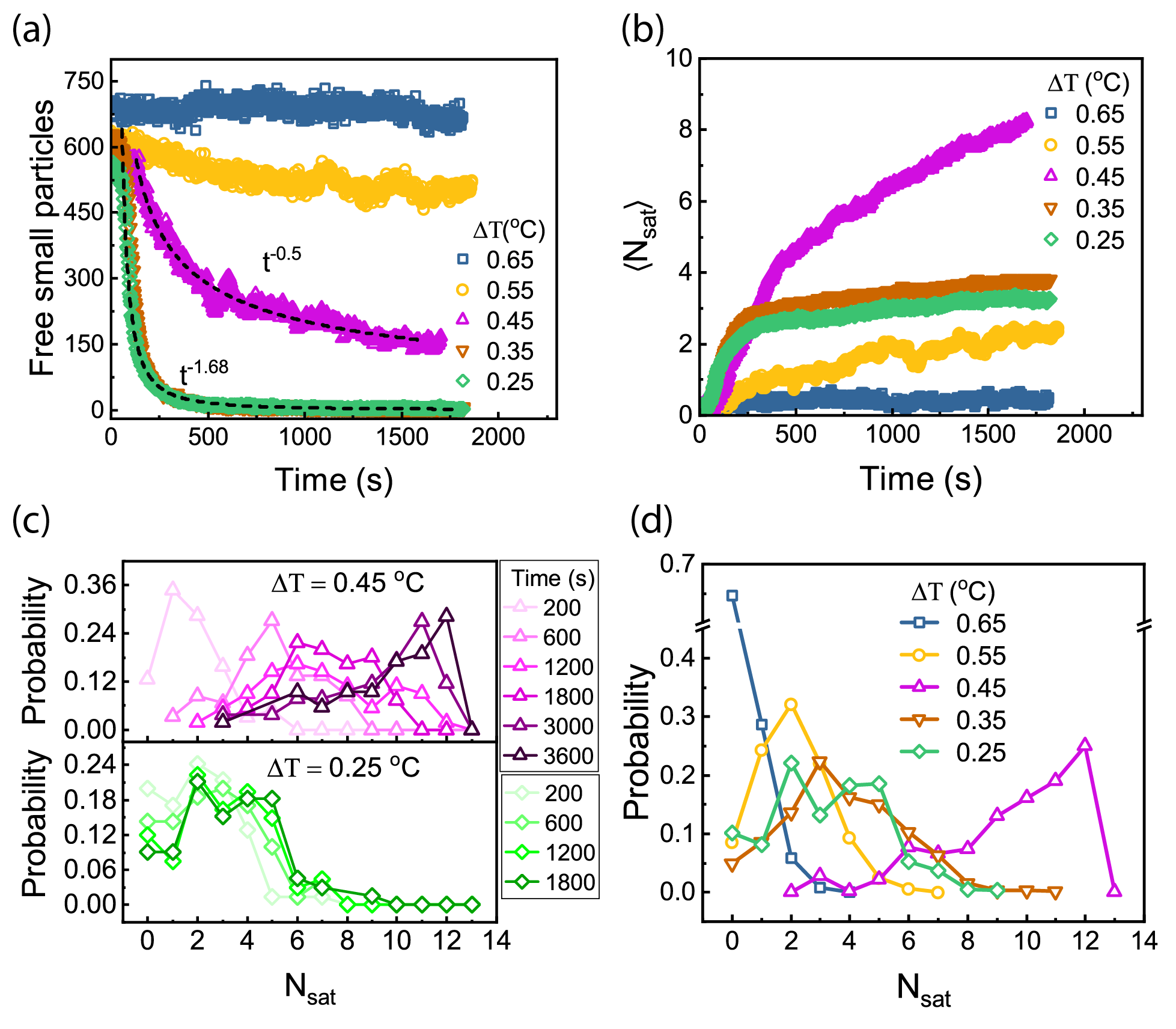}
    \caption{ (a,b) Temporal evolution of (a) the number of free small particles and (b) average number of satellite particles in the first coordination shell around large particles at different $\Delta T$. (c)Time-dependent probability distributions of first-shell occupancy at $\Delta T =  0.45$ and $0.25~^\circ\mathrm{C}$. (d) Probability distribution of the maximum first-shell occupancy measured in the final frame for all \(\Delta T\).}
    \label{fig:Satellite_Paticle}
\end{figure}

\begin{figure*}[hbt!]
    \centering
    \includegraphics[width=\textwidth]{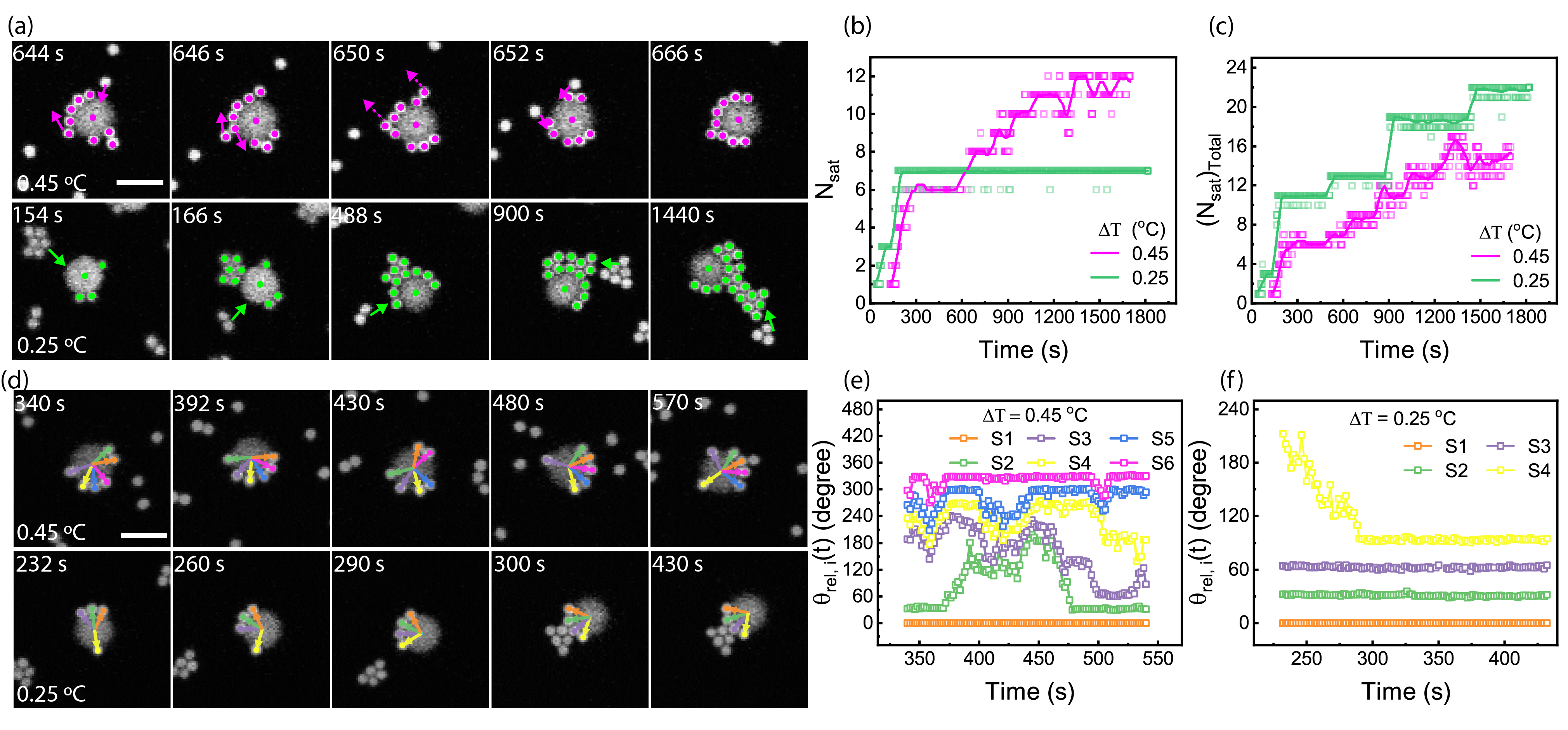}
    \caption{
    (a) Time-resolved CLSM images with detected particle positions during the growth of single binary clusters at (top row, magenta) \(\Delta T = 0.45~^\circ\mathrm{C}\) and (bottom row, green) \(\Delta T = 0.25~^\circ\mathrm{C}\). Solid and dashed arrows indicate the approach and detachment of particles or sub-clusters, respectively. 
    (b) Number of small particles in the first coordination shell, \(N_{sat}\), and (c) total number of attached outer small particles, \((N_{sat})_{total}\), as a function of time. 
    \rev{(d) Time-resolved CLSM images showing the bond angles of small satellite particles relative to the reference satellite S1 (highlighted in orange). (e,f) Relative bond angles, $\theta_{\mathrm{rel, i}}(t)$, over time. (e) Large fluctuations at \(\Delta T = 0.45~^\circ\mathrm{C}\) indicate reversible bonding between the particles that allows reorganization process within the cluster. (f) nearly constant relative bond angles at \(\Delta T = 0.25~^\circ\mathrm{C}\) reflect irreversible bonding and suppressed configurational dynamics. Scale bars are (a) $5~\mu m$ and (d) $5~\mu m$. }}
    \label{fig:cluster_growth}
\end{figure*}

The observed transition in the final structure of the binary clusters indicates that a different assembly mechanisms must be present in each case. To understand how the cluster growth dynamics and distribution of small particles during assembly change when the pair-interactions change, we investigated the change in number of free small particles and the small particle occupancy of the first coordination shell surrounding the large particles over time.
Fig.~\ref{fig:Satellite_Paticle}(a)  shows the temporal evolution of the number of free small particles at different $\Delta T$ and thus different dominant pair-interactions.
At \(\Delta T = 0.55~^\circ\mathrm{C}\), where L-S and S-S attractions are weak, the number of free small particles remains nearly constant as expected. 
Introducing the L-S pair-interaction, \(\Delta T = 0.45~^\circ\mathrm{C}\), results in that the free-particle population decreases over time following a power-law decay with an exponent $\sim0.5$, reflecting gradual incorporation into clusters. 
At \(\Delta T \leq 0.35~^\circ\mathrm{C}\), where S-S attractions also become significant, the free-particle population decreases more rapidly, with an approximate power-law decay exponent of $\sim1.68$, indicating faster aggregation.
Fig.~\ref{fig:Satellite_Paticle}(b) shows the average number of small satellite particles, \(\langle N_{sat}\rangle\), in the first coordination shell of each large particle and reveals a non-monotonic dependence on attraction strength. For the dominant L-S pair-interaction, where the core-shell clusters are formed, the highest coverage is observed. Here, \(\langle N_{sat}\rangle\) increases continuously over time and reaches \(\langle N_{sat}\rangle\) $\approx 8$ at the end of the experiment.
In contrast, for dominant S-S pair-interactions, the average shell occupancy is substantially lower, and after quick increase it can be observed to plateau, reaching \(\langle N_{sat}\rangle\) \(\sim 3\) satellites for \(\Delta T = 0.25~^\circ\mathrm{C}\).
The probability distribution of the maximum shell occupancy over time is shown in Fig.~\ref{fig:Satellite_Paticle}(c) for \(\Delta T = 0.45\) and \(0.25~^\circ\mathrm{C}\), where L-S and S-S pair-interactions are dominant, respectively.  
The probabilities show that only for dominant L-S pair-interactions, the number of small particles in the first shell increases gradually towards maximum of 12 satellites particles, consistent with the geometric packing limit (see Experimental section). We do note that here two later time steps (3000s and 3600s) have been included that lie beyond the earlier reported maximum time of 1700s to exclude a temperature instability (full time evolution of \(\langle N_{sat}\rangle\) for \(\Delta T = 0.45~^\circ\mathrm{C}\) is shown in  Fig. S4 of the Supplementary Information). However, these data points confirm that the maximum packing of shell particles can be reached in the system when given enough time. For dominant S-S pair-interactions the probability distribution appear to be constant, indicating little rearrangements after assembly, consistent with kinetic trapping. 
The probability distribution of the maximum first-shell occupancy measured in the final frame for all \(\Delta T\) is shown in Fig.~\ref{fig:Satellite_Paticle}(d). The distributions show that the maximum first-shell occupancy of $N_{sat}= 12$ is reached when L-S interactions are significant and S-S attractions remain weak, whereas, when S-S attractions become significant, the maximum occupancy does not exceed $N_{sat}\approx 8$ within the experimental observation time.

The observation that compact core-shell structures form for dominant L-S pair-interactions, closely resemble colloidal molecules reported in binary colloidal systems assembled using DNA-mediated interactions \cite{chakraborty_self-assembly_2022, schade_tetrahedral_2013}, external fields \cite{khalil_binary_2012, heatley_colloidal_2017, erb_magnetic_2009}, or active-passive particle mixtures \cite{wang_active_2020, schmidt_light-controlled_2019}. In these studies also the accessible cluster geometries were primarily determined by the particle size ratio. In contrast to our study however, in these systems the pair-interactions were designed such that only contain L-S interactions were present. 
Our results show that for a binary system where multiple pair-interactions are present, tuning the relative strengths of the L-L, L-S, and S-S interactions can also be used to direct the assembly pathway towards the formation of colloidal molecules.

To understand how the change in strength of the small particle L-S and S-S pair interactions control the switch from the core-shell clusters to a branched network, we investigated the mechanism responsible for this behavior. For this we examined the growth dynamics of individual binary clusters and quantify the extent of structural reorganization of the small particles during cluster assembly.
Fig.~\ref{fig:cluster_growth}(a) and Supplementary Movie S3 and Movie S4 shows representative CLSM images together with the detected particle positions for the growth of an individual cluster at \(\Delta T = 0.45~^\circ\mathrm{C}\) and \(0.25~^\circ\mathrm{C}\). Here two distinctly different small particle dynamics and growth of the clusters can be observed, where arrows indicate attachment and detachment of the small particles or small particle clusters. \rev{As discussed above, these dynamics correspond to attachment, detachment, and rearrangement of the satellite particles within the wall-confined first coordination shell.}
To quantify the growth in more detail, we tracked the occupancy of the first coordination shell and the total number of attached small particles over time (Fig.~\ref{fig:cluster_growth}(b,c)). At \(\Delta T = 0.45~^\circ\mathrm{C}\), the first-shell occupancy of the small particles around the large particles fluctuates strongly due to reversible bonding. In addition, continuous rearrangement occur and progressively denser packing until the shell reaches its maximum occupancy of 12 particles. At \(\Delta T = 0.25~^\circ\mathrm{C}\), the occupancy increases stepwise and saturates at lower values. Because rearrangements are strongly suppressed, incoming particles attach to existing satellites rather than filling vacant shell positions, resulting in loosely packed, branched structures despite continued cluster growth.
Next, we examined the local satellite particle dynamics to quantify structural rearrangements within the first coordination shell. For this we measured the relative bond angles of neighboring satellite particles relative to a reference satellite particle (Fig.~\ref{fig:cluster_growth}(d-f)). At $\Delta T = 0.45~^\circ\mathrm{C}$, relative bond angles fluctuate substantially \rev{while the first coordination shell remains incompletely occupied, indicating persistent rearrangements and rotational motion of the small particles within the available shell positions (Supplementary Movie S5). As the shell approaches full occupancy, the angular fluctuations decrease as the available configurational space becomes increasingly restricted by neighboring satellite particles. In contrast at $\Delta T = 0.25~^\circ\mathrm{C}$, relative bond angles remain nearly constant even at lower shell occupancy, indicating that the reduced configurational dynamics cannot be explained solely by the limited available space in a densely occupied shell, but instead reflects the suppression of rearrangements by the stronger particle associations (Supplementary Movie S6).
To further verify this behavior, we analyzed additional individual binary clusters and relative angular variations of satellite particles at both temperatures (Fig.~S5(a,b)).}
The reduced bond-angle fluctuations when S-S pair interactions are significant indicate that structural rearrangements become increasingly suppressed after bond formation. Similar behavior has been reported in short-range attractive colloidal systems, where strong attractions limit restructuring and promote kinetically constrained assembly pathways \cite{hecht_kinetically_2016}.
We conclude that when the L-S pair interactions are present the clusters can still undergo frequent attachment and detachment of small particles that are highly dynamic and rearrange often and ultimately leads to the formation of compact core-shell structures with a nearly symmetric distribution of small particles around the central large particle. In contrast, when the S-S pair-interactions are also present small particles cluster form first and then attach to large particles, while at the same time L-S pair-interactions increase in strength causing small particles remain bonded to the large particles once attached. It is the combination of these two processes that results in kinetic arrest and the growth of anisotropic and branched clusters. 
This behavior is consistent with the non-monotonic shell occupancy observed in Fig.~\ref{fig:Satellite_Paticle}, suggesting that efficient shell formation requires attractions strong enough to drive particle incorporation, but not so strong that particle mobility and local rearrangements are strongly suppressed \cite{kozina_crystallization_2014}.
Together, these observations demonstrate that shell filling is controlled by the dynamics and rearrangements of the small particles: dominant L-S pair-interactions permit structural reorganization and efficient packing of the first coordination shell, whereas dominant S-S pair-interactions suppress rearrangements and promote branched growth.
\rev{These two regimes show qualitative similarities to reaction-limited and diffusion-limited aggregation. The reversible attachment and rearrangement observed at \(\Delta T=0.45^\circ{\rm C}\) resembles reaction-limited aggregation, whereas the persistent bonding and branched growth at \(\Delta T\le0.35^\circ{\rm C}\) resembles diffusion-limited aggregation. We note, however, that the present multicomponent, wall-confined system is not a classical RLCA/DLCA system.}

\subsection{Bulk assembly process}

\begin{figure}
    \centering
    \includegraphics[width=\linewidth]{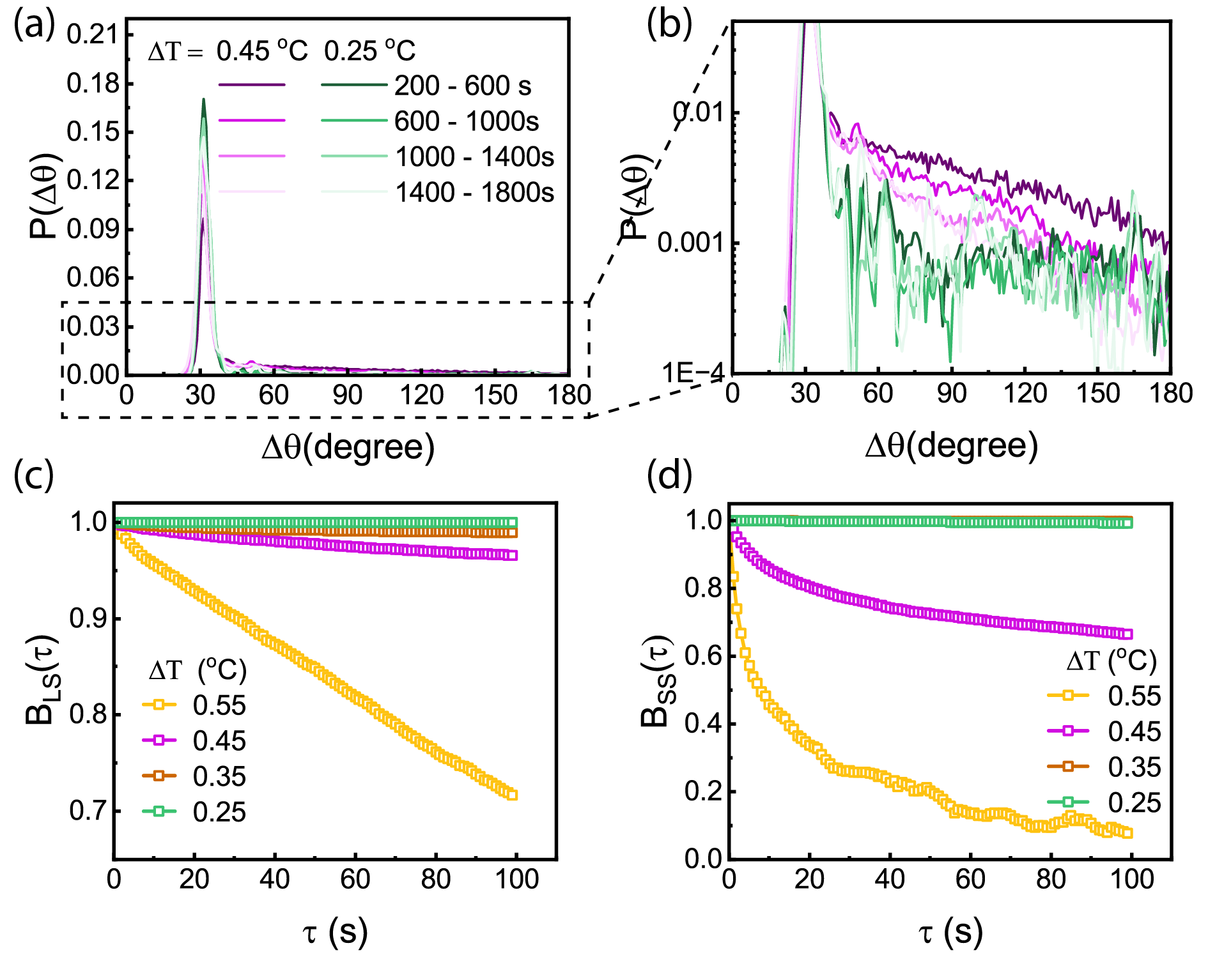}
    \caption{\rev{(a)Probability distributions $P(\Delta\theta)$ of angular separations between consecutive small satellite particles around large particles at $\Delta T=0.45\,^\circ\mathrm{C}$ and $0.25\,^\circ\mathrm{C}$ over successive $400$s intervals from 200 to 1800~s. Distributions include all analyzed binary clusters and frames within each interval. (b) The same distributions on a logarithmic scale, highlighting the broader tail at $\Delta T=0.45\,^\circ\mathrm{C}$, indicating greater heterogeneity in satellite angular arrangements.}
    (c,d) Bond correlation, \(B(\tau)\), as a function of delay time, \(\tau\) for (c) large-small and (d) small-small pair.}
    \label{fig:Bond_corr}
\end{figure}

To determine whether these single-cluster dynamics are representative of the bulk assembly process, we investigated the dynamics of all observed binary colloid clusters. \rev{Fig.~\ref{fig:Bond_corr}(a,b) shows the distributions of angular separations between consecutive satellite particles in the first coordination shell of the large particles, over successive time intervals of $\Delta t = 400$s between 200 to 1800~s. The distribution at $\Delta T = 0.45~^\circ\mathrm{C}$ shows a broader tail toward large angular separations than at $\Delta T = 0.25~^\circ\mathrm{C}$, consistent with greater configurational mobility of the smaller particles at the weaker attraction. Next, we addressed the overall satellite dynamics further, by calculating }the bond correlation function, given by \(B(\tau) = \sum_{\substack{i, j = 1 \\ i \neq j}}^{N} \langle b_{ij}(t) b_{ij}(t+\tau) \rangle\), where $b_{ij}(t)$ is equal to 1 if particles $i$ and $j$ are bonded at time $t$, and 0 otherwise. \rev{Thus, \(B(\tau)\) measures whether a particle pair that is bonded at time \(t\) also satisfies the bond criterion at time \(t+\tau\). The bond state at intermediate times is not explicitly tracked, and therefore a bond that breaks and reforms within the interval may still contribute to \(B(\tau)\).} To evaluate this quantity, we construct an $N \times N$ matrix $\mathbf{b}(t)$, whose elements are given by $b_{ij}(t)$ for all particle pairs \cite{navas_impact_2024}.
As the final cluster structure is primarily governed by the L-S and S-S interactions, we focus on the corresponding bond correlations for \(0.25~^\circ\mathrm{C}\leq\Delta T\leq 0.55~^\circ\mathrm{C}\) shown in Fig.~\ref{fig:Bond_corr}.
Fig.~\ref{fig:Bond_corr}(c) presents the bond correlation functions for L-S pairs. At \(\Delta T=0.55~^\circ\mathrm{C}\), \(B_{LS}(\tau)\) decays rapidly, indicating weak and short-lived associations between large and small particles. At \(\Delta T=0.45~^\circ\mathrm{C}\), the decay becomes significantly slower, demonstrating that L-S bonds persist for longer times while remaining reversible. In contrast, for \(\Delta T\leq0.35~^\circ\mathrm{C}\), \(B_{LS}(\tau)\) remains nearly constant over the experimental timescale, indicating that bond-breaking events become strongly suppressed.
A similar trend is observed for S-S pairs (Fig.~\ref{fig:Bond_corr}(d)). At \(\Delta T = 0.55\) and \(0.45~^\circ\mathrm{C}\), \(B_{SS}(\tau)\) exhibits measurable decay, showing that bonds between small particles remain dynamic. As the attraction strength increases further, the bond correlation becomes essentially time-independent, indicating the formation of long-lived S-S bonds that rarely break during the observation window.
\rev{At long lag times, the calculation is additionally biased toward particle pairs that can be tracked for sufficiently long times, which preferentially selects less mobile particles and particles within clusters. Together with statistical fluctuations from averaging over different time origins, this can result in the slight non-monotonic behavior observed for \(B_{SS}(\tau)\) at \(\Delta T=0.55^\circ{\rm C}\).}
These results confirm a progressive transition from reversible to effectively irreversible bonding when introducing the S-S pair-interactions. The increasing bond persistence is consistent with the reduced bond-angle fluctuations and suppressed structural rearrangements observed under these conditions. Consequently, particles become trapped shortly after attachment, providing a microscopic explanation for the reduced shell occupancy and the emergence of branched cluster morphologies.

\begin{figure}
    \centering
    \includegraphics[width=\linewidth]{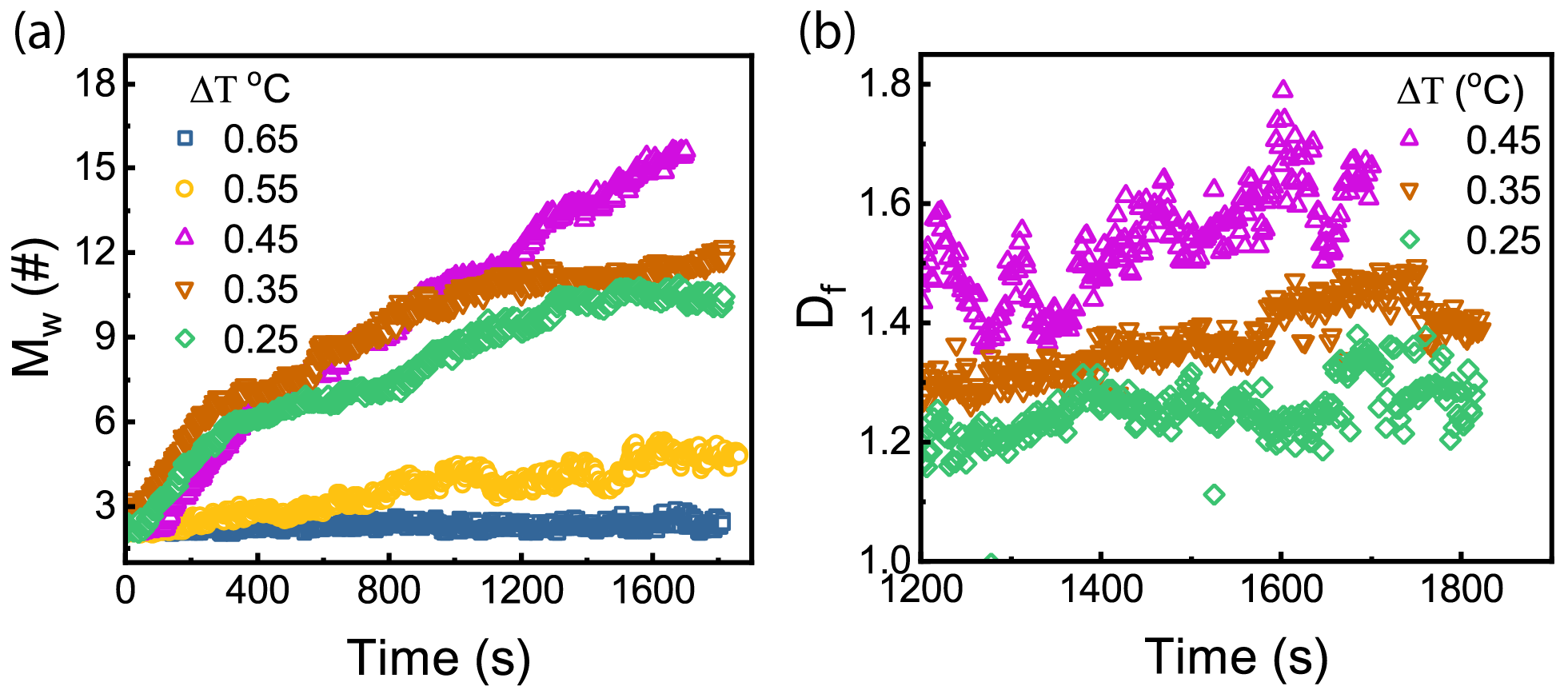}
    \caption{Temporal evolution of (a) mean cluster size, \(M_w\) for all \(\Delta T\) and (b) fractal dimension \(D_f\) for \(\Delta T = 0.45, 0.35,\) and \(0.25~^\circ\mathrm{C}\). }
    \label{fig:FractalDimension}
\end{figure}

Finally, we investigated how these microscopic mechanism affect the evolution of the overall cluster size and the large-scale morphology of the assembled structures.
Fig.~\ref{fig:FractalDimension}(a) shows the temporal evolution of the mean binary cluster size, \(M_w\) for all investigated $\Delta T$ and thus combinations of pair-interactions. The analysis was restricted to binary clusters in order to specifically probe the influence of inter-species interactions on cluster growth. The largest binary clusters are observed when the L-S pair interactions dominate (\(\Delta T = 0.45~^\circ\mathrm{C}\)) and promote the incorporation of small particles around the large-particle cores. As the S-S pair-interactions are introduced and become stronger the mean binary cluster size decreases despite the overall enhancement of particle aggregation. Owing to their higher diffusivity, small particles rapidly assemble into small particle clusters, resulting in a large population of isolated small-particle clusters within the field of view. Consequently, fewer small particles remain available for incorporation into binary clusters.
To quantify the resulting cluster morphology, we determine the fractal dimension, \(D_f\), from the scaling relation between the cluster size \(M\) and the radius of gyration \(R_g\), \(R_g \sim M^{1/D_f}\). 
Fig.~\ref{fig:FractalDimension}(b) shows the temporal evolution of \(D_f\) for \(\Delta T = 0.45\), \(0.35\), and \(0.25~^\circ\mathrm{C}\).
The fractal dimension decreases systematically upon going from L-L and L-S pair-interactions to increasing S-S pair-interactions, indicating a transition from relatively compact clusters to increasingly ramified structures. The largest \(D_f\) values are observed at \(\Delta T = 0.45~^\circ\mathrm{C}\), where the first coordination shell achieves its highest occupancy and particles retain the ability to reorganize during growth. In contrast, the lowest \(D_f\) values occur at \(\Delta T = 0.25~^\circ\mathrm{C}\), where growth proceeds through irreversible particle attachment and branch formation.
This sensitivity of cluster morphology to the assembly mechanism is consistent with observations in multi-component colloidal systems, where the final structure is governed not only by the strength of particle interactions but also by how different particle species participate during growth \cite{mugnai_interspecies_2025, hecht_kinetically_2016}. 
Taken together, the evolution of both \(M_w\) and \(D_f\) demonstrates that the presence of multiple pair interactions influences not only the extent of aggregation but also the architecture of the resulting assemblies. 

\section{Conclusion}
We have shown by observing the nucleation and growth of binary colloid clusters of large and small colloids induced by critical Casimir forces that the strength of the small particle pair-interactions, i.e. L-S and S-S, has a distinct effect on the assembly mechanism. By controlling the temperature, we systematically changed the strength of the L-L, L-S, S-S pair-interactions. Our experiments show that when besides the L-L pair-interactions, the L-S pair-interactions are present, the formation of compact core-shell structures occurs of small satellite particles around the large particles, also known as binary molecules. In contrast, when additional S-S pair-interactions are present, extended satellite-particle networks are formed that result in branched binary clusters. We further showed via single-cluster growth measurements and bond-correlation analysis that these structural transitions are governed by strength of the L-S and S-S pair-interactions that change the dynamics and degree of structural reorganization of the small particles during the cluster assembly. Dominant L-S pair-interaction permits repeated bond breaking and rearrangement events of the small particles, allowing defects to anneal out and consequently a progressive filling of the first coordination shell around a large particle. In contrast, dominant S-S pair-interactions suppress these rearrangements, leading to kinetically arrested growth in which small and large particles become trapped in non-equilibrium configurations shortly after attachment. 
Our results establish a direct link between the interaction strength of pair-interactions and the assembly mechanisms in binary colloid systems and shed new light onto how these control the local particle organization, structural reorganization, and morphology. These insights are important for how  small particle interactions, dynamics and assembly pathways influence the formation of space-spanning binary colloidal networks and gels.
The understanding of how tuning the pair-interactions influence particle association and configurational rearrangement \rev{shows that compact binary clusters can be promoted by maintaining sufficient bond reversibility for local reorganization, whereas stronger S-S interactions suppress rearrangement and favor kinetically arrested branched structures.}

\section*{Author Contributions}
Conceptualization – DKM, JMM; 
formal analysis – DKM, TWJV, BES; 
investigation – DKM, BES; 
methodology –DKM, BES, JMM; 
supervision – DKM, JMM, WGE; 
visualization – DKM, BES, TWJV; 
writing (original draft) – DKM, JMM; 
writing (review and editing) – DKM, JMM, WGE. 
All authors have read and approved the final text of the paper.

\section*{Conflicts of interest}
The authors declare no conflict of interest.

\section*{Data availability}
Data is available from the authors upon reasonable request. 

\section*{Acknowledgements}
The authors thank Max van Donkelaar for his contributions to this work as part of his Bachelor's thesis. 
This work was supported by the Netherlands Organization for Scientific Research (NWO) grant VI.Vidi.223.126.

\section*{References}
\makeatletter
\renewcommand{\@biblabel}[1]{#1}
\providecommand{\bibnumfmt}[1]{#1}
\renewcommand{\bibnumfmt}[1]{#1}
\makeatother

\bibliographystyle{rsc} 
\bibliography{references} 

\end{document}